\documentclass[prb,aps,twocolumn,showpacs]{revtex4}

\usepackage{epsfig}

\begin{document}

\title{Novel magnetic field tuning of quantum spin excitations in a weakly coupled $S=$1/2 Heisenberg spin chain as seen from NMR}

\author{Long Ma$^{1}$}
\author{Z. Wang$^{1}$}
\author{L. Hu$^{1}$}
\email{hulin@hmfl.ac.cn}
\author{Z. Qu$^{1}$}
\email{zhequ@hmfl.ac.cn}
\author{N. Hao$^{1}$}
\author{Li Pi$^{1,2}$}
\email{pili@ustc.edu.cn}

\affiliation{$^{1}$Anhui Province Key Laboratory of Condensed Matter Physics at Extreme
Conditions, High Magnetic Field Laboratory, Chinese Academy of Sciences, Hefei
230031, China\\
$^{2}$ Hefei National Laboratory for Physical Sciences at the Microscale, University of Science and Technology of China, Hefei 230026, China}

\date{\today}

\pacs{75.10.Pq, 75.40.Gb, 76.60.-k}

\begin{abstract}
We report our NMR study of the spin excitations in the quasi-one dimensional (1D) $S=1/2$ quantum magnet CH$_3$NH$_3$Cu(HCOO)$_3$ lying in the 1D-3D dimensional
crossover regime with Dzyaloshinskii-Moriya(DM) interactions. Above $T_N$, the spinon excitation is observed from the constant $1/T_1$ at low temperatures contributed from the staggered spin susceptibility. At low temperatures well below $T_N$, the $1/T_1$ begins to flatten out under weak magnetic fields.
With the increasing field intensity, $1/T_1$ tends to show a power-law temperature dependence gradually, with the index increasing from zero to $\sim3$, and finally $\sim5$, which are the respective typical characteristic for the dominating two-magnon Raman process and three-magnon scattering contributions to the nuclear relaxation in a conventional 3D magnet. A possible physical mechanism for this novel magnetic field tuning of quantum spin excitations related with the enhanced effective staggered field created by the DM interactions under magnetic field is discussed.

\end{abstract}

\maketitle

In low-dimensional (low-D) magnets, the enhanced quantum fluctuations lead to novel states of matter, exotic elementary excitations with integer or fractional
quantum numbers, as well as fascinating critical phenomena\cite{Zapf_RMP, Han_nature, Punk_NP, Piazza_NP}. Studies on the quantum-mechanical effects in low-D
magnets have triggered enormous and unfailing research interest in the condensed matter society, and have profound influences on the understanding of high-$T_C$
superconductivity, fractional quantum Hall effect, and other strongly-correlated electron systems\cite{Anderson_Science_235_1196, Kivelson_PRB_35_8865,
Raghu_PRL_100_156401, Kane_PRL_95_226801}. Among others, a prototypical example is the one-dimensional $S=1/2$ Heisenberg antiferromagnetic spin chain (HAFC),
where the strong quantum fluctuations prohibit any possible N\'{e}el order even at zero temperature limit\cite{Mermin_PRL_17_1133}. This spin system has received
long-lived research interest since it's analytically solvable\cite{Bethe_ZPhys_71_205}, provides a rare opportunity for rigorous comparison between the experimental
 results and theoretical approaches\cite{Tennant_PRB_52_13368, Hammar_PRB_59_1008}, and gives a valuable test of the reliability of the theoretical approximation.

The $S=1/2$ HAFC shows a gapless spin excitation spectrum with a large area of continuum, and the excitation quasiparticle is fractionalized
with $S=1/2$ (spinon)\cite{Faddeev_PLA_85_375, Haldane_PRL_66_1529,Mourigal_NP_9_435}. When the inter-chain coupling is included, 3D long range N\'{e}el order
will be restored at sufficiently low temperatures. The effective staggered field resulting from the inter-chain coupling confines pairs of the
quantum spinons into classical magnons, the Goldstone mode of the translational symmetry breaking. From the spin dynamics, the spin excitation
spectrum of the quasi-1D spin chain system is dominated by the sharp dispersive spin waves at the low energy region and the continuum spectrum formed
by multi-spinon excitations at higher energies\cite{Lake_NM_4_329}. The longitudinal mode of the spin wave excitation is proposed theoretically based on mean-field or
Random phase approximations (RPA)\cite{Schulz_PRL_77_2790, Essler_PRB_56_11001}. This is substantially different with the spin excitations in the conventional 3D antiferromagnets, which is mainly contributed by the well-defined transverse spin waves. By tuning the inter-chain coupling, the dimensional crossover from the critical quantum disordered 1D spin system to the classical 3D magnets occurs, which is important for understanding the semiclassical behavior induced by underlying quantum fluctuations in the materials in the real 3D world.

The quasi-1D HAFC is realized in several compounds with very different magnetic coupling strength, KCuF$_3$\cite{Satija_PRB_21_2001, Hutchings_PR_188_919, Lake_NM_4_329}, Sr$_2$CuO$_3$ and SrCuO$_2$\cite{Kojima_PRL_78_1787, Matsuda_PRB_55_11953, Zaliznyak_PRL_83_5370}, BaCu$_2$Si$_2$O$_7$\cite{Zheludev_PRL_85_4799, Tsukada_PRB_60_6601, Zheludev_PRL_89_197205} et al. The dimensionality can be measured by the ratio between the intra-chain coupling $J$ and the N\'{e}el ordered temperature $T_N$, also the ordered moment $m_0$, as the $T_N$ is roughly proportional to the inter-chain coupling strength and the quantum fluctuations can strongly suppress the static moment in the ordered state. For KCuF$_3$, $J=17$ meV, $T_N=39$ K ($J_{\perp}/J\sim0.084$), and $m_0=0.5 \mu_B$\cite{Satija_PRB_21_2001, Hutchings_PR_188_919}, which is close to the 3D spin system. The nearly ideal 1D-character is realized in Sr$_2$CuO$_3$ and SrCuO$_2$, with a strong intra-chain exchange interaction energy $J$ of $\sim 190$ meV and $\sim 181$ meV , a low $T_N$ of 5.4 K and 2 K ($J_{\perp}/J\sim7\times10^{-4}$ and $2.5\times10^{-4}$), and a very small ordered moment both with the order of $<0.1\mu_B$(close to or below the detecting limit), respectively\cite{Kojima_PRL_78_1787, Matsuda_PRB_55_11953, Zaliznyak_PRL_83_5370}. The BaCu$_2$Si$_2$O$_7$ lies on the dimensional crossover regime, where the $J$ equals to 24.1 meV and the antiferromagnetic transition occurs at $T_N=9.2$ K ($J_{\perp}/J\sim0.011$) with a static ordered moment of $0.15 \mu_B$ (Refs.~[\onlinecite{Zheludev_PRL_85_4799, Tsukada_PRB_60_6601, Zheludev_PRL_89_197205}] and therein). The novel longitudinal mode is evident by inelastic neutron scattering
with a reduced life time due to the decay into two transverse spin waves in KCuF$_3$\cite{Lake_PRL_85_832, Lake_PRB_71_134412}, consistent with the theoretical predictions. The BaCu$_2$Si$_2$O$_7$ spin system locates on the 1D to 3D crossover regime with an enhanced quantum fluctuations, thus the longitudinal mode should be stronger. Surprisingly, a single broad continuum is observed in the longitudinal spin excitation spectrum, indicating the absence of the longitudinal mode\cite{Zheludev_PRL_89_197205, Zheludev_PRB_67_134406}. This contradiction should be attributed to the ignorance of the correlation effects in the mean-field and RPA approach. Unusual excitations is observed in Sr$_2$CuO$_3$ recently by electron spin resonance\cite{Sergeicheva_PRB_95_020411}, which may be related to interactions between the Goldstone magnons and the amplitude fluctuations of the order parameter. As a result, the quantum correlation effects should play an important role in the dominance of the spin excitations when the spin system moves closer to the quantum critical point. According to our knowledge, satisfactory theory describing the spin systems near the 1D-3D dimensional crossover region is still highly lacked.

Nuclear magnetic resonance (NMR), mainly detecting the on-site hyperfine field and its fluctuations resulting from the electron spins via the hyperfine coupling,
is a powerful local probe accessing the magnetic susceptibility, magnetic order as well as spin dynamics near the quantum critical point in weakly coupled
quasi-1D antiferromagnets. Several NMR results concerning the Luttinger physics in quasi-1D quantum antiferromagnets tuning by
the applied magnetic field are reported\cite{Jeong_PRL_118_167206,Klanjsek_PRL_101_137207,Mukhopadhyay_PRL_109_177206,Casola_PRB_86_165111}, important for the relative research on quantum criticality behaviors. In this paper, we report our NMR study on the CH$_3$NH$_3$Cu(HCOO)$_3$ (Cu-MOF) quantum magnet lying on the 1D-3D crossover regime. The $G$-type antiferromagnetic order with spin canting is proposed from the spectral analysis. Above $T_N$, free spinons are observed from the constant $1/T_1$ contributed from the staggered spin susceptibility. Below $T_N$, the $1/T_1$ begins to flatten out at low temperatures at the weak magnetic field side.
By increasing the field intensity, the $1/T_1$ gradually shows a power-law temperature dependence at low temperatures with the index of $\sim3$ to $\sim5$, corresponding to the two-magnon dominant and three-magnon dominant contributions to the nuclear relaxation in a classical 3D magnet.
This novel magnetic field tuning of the quantum magnetic fluctuations in the ordered state is possibly related with the enhanced effective staggered field generated by the Dzyaloshinskii-Moriya(DM) interactions under magnetic field. This newly discovered phenomena should be fresh physical inputs for constructing proper theoretical models.

\begin{figure}
\includegraphics[width=8cm, height=8cm]{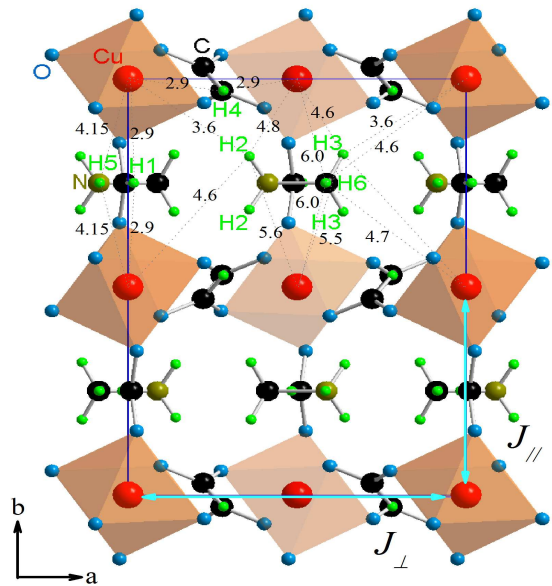}
\caption{\label{struc1}(color online) The crystal structure of CH$_3$NH$_3$Cu(HCOO)$_3$ viewed from $c$-axis. The CuO$_6$ octahedron with obvious Jahn-Teller distortion
is shown by the orange polyhedra. The crystallographically inequivalent positions of hydrogen is marked with H$_1$, H$_2$ to H$_6$, whose distance with the nearest
Cu$^{2+}$ neighbours is also shown by the dotted line and numbers in angstrom.
}
\end{figure}

High quality Cu-MOF single crystals are synthesized under the solvothermal
conditions\cite{Pato_JMCC_4_11164}. Single crystals with typical dimensions of $2\times2\times1.5$ mm$^3$ are choosen for our NMR study.
Our NMR measurements are conducted on the $^1$H nuclei($\gamma_n=42.5759$ MHz/T, $I=1/2$) with a phase-coherent NMR spectrometer.
The spectra are obtained by summing up the spin-echo intensities as a function of frequency.
The spin-lattice relaxation rates are measured by a conventional inversion-recovery method, and fitting the nuclear magnetization to the standard
relaxation curve for $I=1/2$. All the data are collected during the warming-up process for self-consistence.

The Cu-MOF crystalizes in a orthorhombic structure with space group $Pnma$ ($Z=4$). Fig.\ref{struc1} shows the detailed structure of Cu-MOF,
with the CuO$_6$ octahedron shown by the orange polyhedra. The structure shows a distorted perovskite-like (with a general chemical formula ABX$_3$) characteristic,
with the A sites occupied with the (CH$_3$NH$_3$)$^+$ cations and (HCOO)$^-$ anions occupying the X sites bridging the magnetic interactions between the Cu$^{2+}$
sitting on the B sites. The $3d^9$ configuration of Cu$^{2+}$ in the cubic crystal field makes the Cu-MOF be a Jahn-Teller active system, leading to local
distortions in the crystal structure and the orbital ordering, which is supported by very different Cu-O bonds' length. Based on the Goodenough-Kanamori-Anderson (GKA) rules\cite{Goodenough_magnetism}, shorter Cu-Cu distance along $b$-axis and more significant antiferromagnetic interactions between the half-filled $d_{x^2-y^2}$ through the [HCOO]$^-$ anions may give much stronger antiferromagnetic coupling along $b$-axis\cite{Pato_JMCC_4_11164}, and a weaker ferromagnetic coupling in the $ac$-plane. Experimentally, both the broad Bonner-Fisher peak at $T\sim 45$ K shown in the temperature dependence of the susceptibility and the obviously anisotropic responds of the magnetization to the applied field confirm the quasi-1D magnetism in Cu-MOF\cite{Pato_JMCC_4_11164}. This mechanism is very similar to what occurs in the model spin system KCuF$_3$ with a similar 3D structure\cite{Hutchings_PR_188_919}. The intra-chain (along $b$-axis)coupling $J$ is estimated to be $\sim 6$ meV from Bonner-Fisher fitting to the susceptibility, and the $T_N$ is $\sim4$ K at low magnetic field\cite{Pato_JMCC_4_11164}. Based on the mean-field theory, the ratio between the inter- and intra-chain coupling is estimated to be $J_{\perp}/J\sim0.02$. Therefore, the spin system in Cu-MOF locates on the 1D-3D crossover regime, between BaCu$_2$Si$_2$O$_7$ and KCuF$_3$.

We conduct our NMR study on Cu-MOF via the $^1$H site, as strong wipe-out effect of $^{63, 65}$Cu nuclei make the signals undetectable due to too fast
spin-spin relaxations on the magnetic sites. Six crystallographically inequivalent positions of hydrogen exist in Cu-MOF, which is marked with H$_1$, H$_2$ to H$_6$
in Fig.\ref{struc1} and the distance between them and their nearest Cu$^{2+}$ neighbours is also shown. In this magnetic insulator, the hyperfine coupling is mainly
contributed from the dipolar interactions between the nuclear spins and electron orbitals and spins, which is very sensitive to the distance between the nuclei and
magnetic ions. Obviously, the H$_1$ and $H_4$ sites from [HCOO]$^-$ anions locate much nearer to Cu$^{2+}$ sites. Additionally, the [HCOO]$^-$ anion bridging the magnetic
interactions, has chemical bonds with Cu$^{2+}$ sites, while the (CH$_3$NH$_3$)$^+$ cations link with the framework via much weaker H-bonds. Thus, the hyperfine coupling of  H$_1$ and $H_4$ sites from [HCOO]$^-$ anions should be much stronger than the other four hydrogen sites.

\begin{figure}
\includegraphics[width=8.5cm, height=8cm]{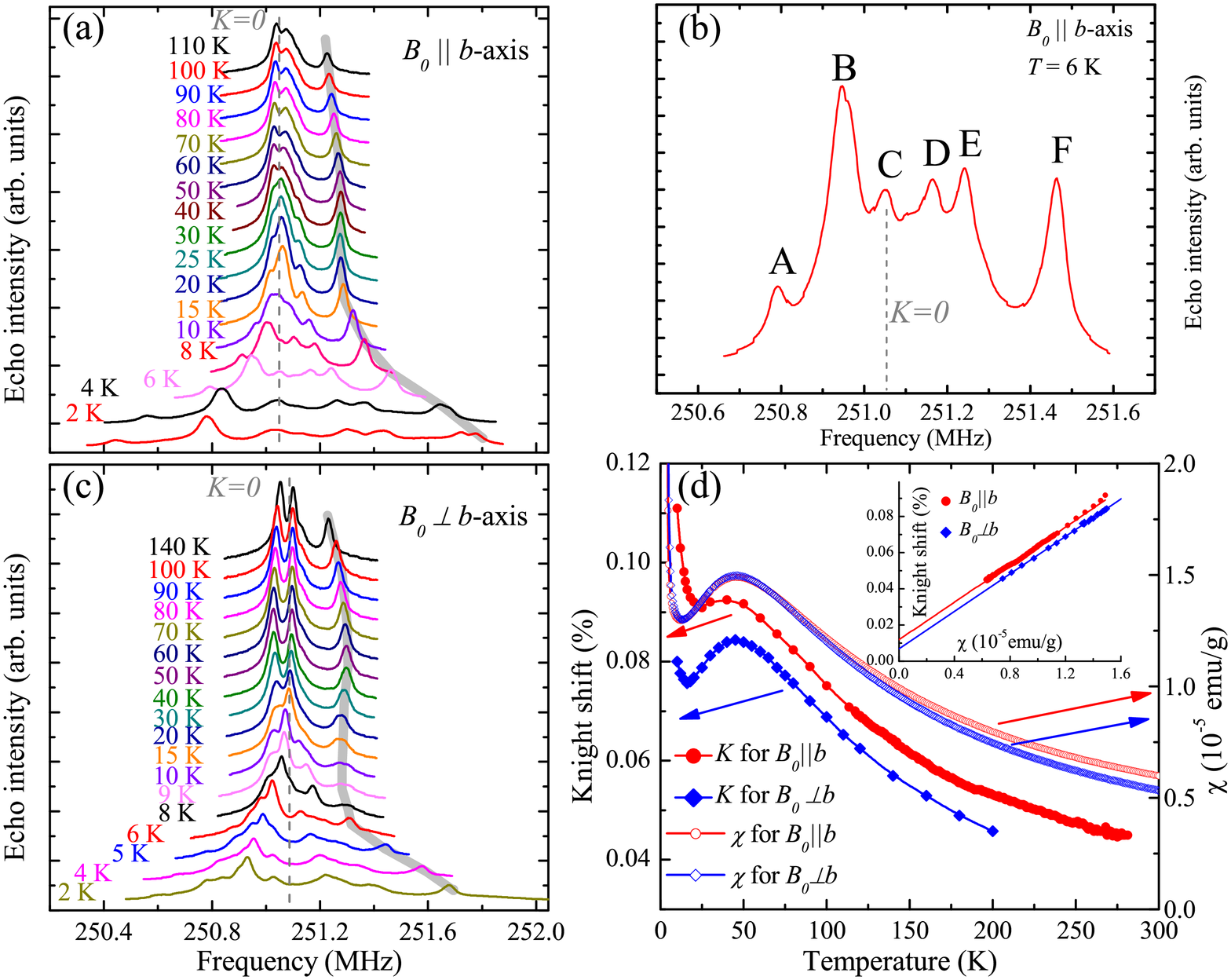}
\caption{\label{spec1}(color online)
(a): The $^1$H NMR spectra under different temperatures with a 5.9 T magnetic field applied along the crystalline $b$-axis. The vertical dashed line show the Larmor
frequency $\gamma_nB_0$. (b): An enlarged version of the spectrum at $T=6$ K in (a). (c): Temperature dependence of $^1$H NMR spectra with $B_0\perp b$-axis.
(d): The Knight shift and dc susceptibility versus temperature for both field directions. For the susceptibility measurement, a field of 200 Oe is applied, and the
data are collected with the sample warming up after a zero field cooling (ZFC) process. Inset: the Knight shift as a function of dc susceptibility with the temperature
as an explicit parameter (See the text).
}
\end{figure}

Typical frequency-swept NMR spectra at various temperatures are shown in Fig.\ref{spec1} (a) and (c) for both field directions. The spectrum is composed by a
single peak at higher frequency and a group of peaks near the Larmor frequency which overlap with each other at high temperatures and separate from others with
the sample cooling down as a result of the increased spin susceptibility. At $T=6$ K above $T_N$, six well defined Lorentz peaks are identified
 (See Fig.\ref{spec1} (b)), corresponding respectively to the six crystallographically inequivalent hydrogen sites. Based on the above discussions of the hyperfine
 coupling shown above, we assign the $A$ and $F$ peaks to the hydrogen in [HCOO]$^-$ anions, and the other four peaks to the hydrogen in (CH$_3$NH$_3$)$^+$ cations.
The Knight shifts (defined as $K=(f-\gamma_n B_0)/(\gamma_n B_0)$, where the $f$ denotes the actual resonance frequency, and $B_0$ the applied magnetic field) of
different peaks show similar temperature dependence, all reflecting the bulk spin susceptibility. We select the peak with the strongest hyperfine coupling ( marked
with the thick gray line in Fig.\ref{spec1} (a) and (c)) to study the magnetic properties in order to gain a better sensitivity.
The temperature dependence of the Knight shift of the F peak and dc susceptibility in the paramagnetic state is shown in Fig.\ref{spec1} (d).
With the sample cooling down, the Knight shifts show a Curie-Weiss upturn at high temperatures, and show a broad peak at $T_{max}\sim45$ K,
consistent with the dc susceptibility. Based on the equation $T_{max}=0.641J_{//}$ expected for $S=1/2$ Heisenberg chains\cite{Klumper_PRL_84_4701},
the intra-chain coupling $J_{//}$ is estimated to be $\sim6$ meV, consistent with previous results. For temperatures below $T\sim 18$ K, the Knight shifts also
show the upturn behavior as seen from the dc susceptibility, indicating an intrinsic enhanced spin susceptibility in Cu-MOF. This behavior is related to the DM interactions in this sample due to the lack of inversion symmetry center, which will be discussed later. As $K=K_{spin}+K_{orbit}=A_{hf}\chi_{spin}+K_{orbit}$, we can extract the diagonal element of the hyperfine coupling constant tensor $A_{hf}$ from the slope of the line by plotting the Knight shift versus dc susceptibility with temperature as an explicit parameter (See Fig.\ref{spec1}(d) inset, also called  Clogston-Jaccarino plot\cite{Clogston_PR_121_1357}). The linear fittings give the hyperfine coupling $A_{hf}^{bb}=1.254$ kOe/$\mu_B$ and $A_{hf}^{aa/cc}=1.252$ kOe/$\mu_B$, which is nearly isotropic.

\begin{figure}
\includegraphics[width=8.5cm, height=8cm]{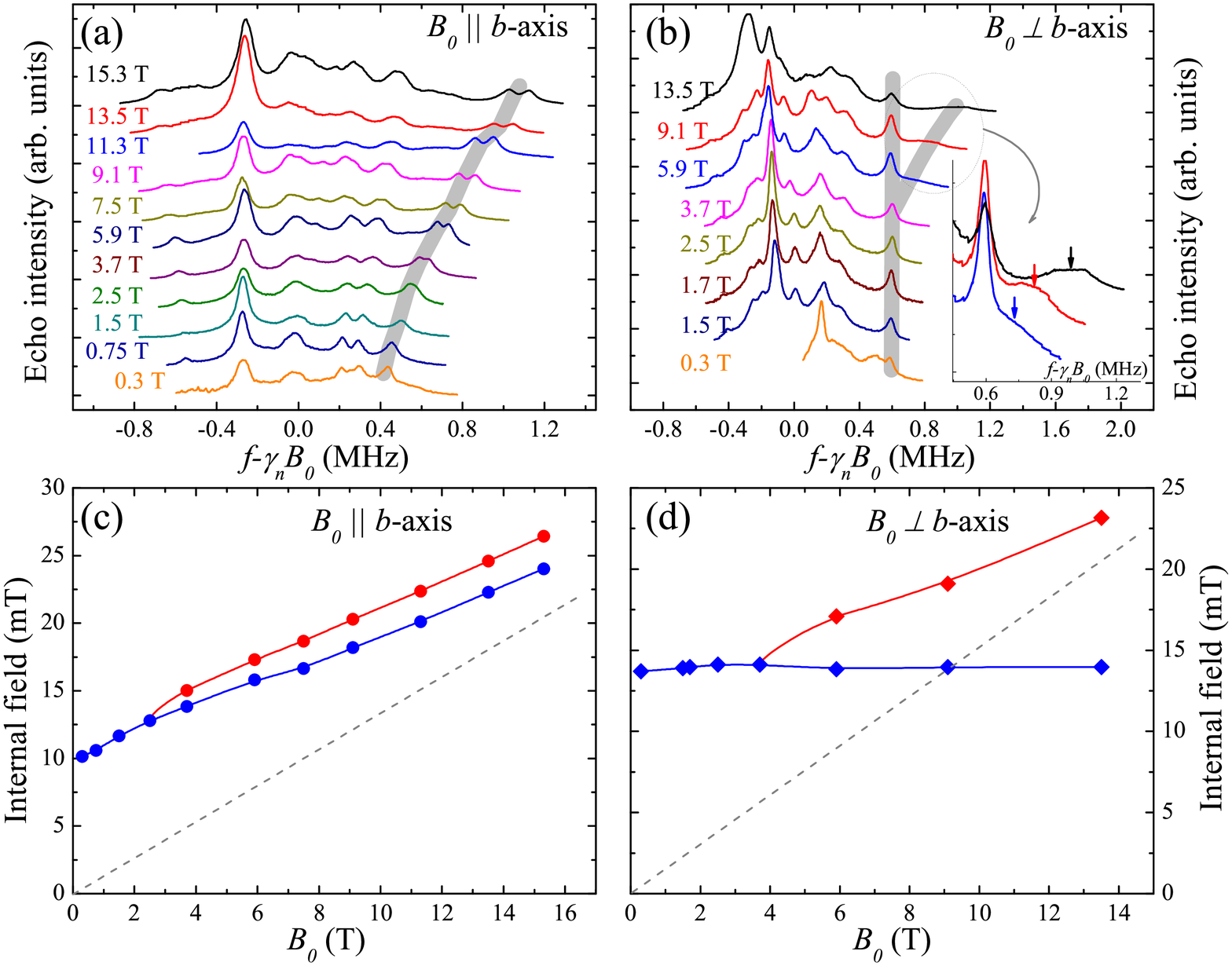}
\caption{\label{spec2}(color online)
  The spectra at $T=2$ K for different field intensities with the $B_0||b$ (a) and $B_0\perp b$-axis (b), respectively.
  The gray lines show the change of the relative frequency shift of the studied signal peak. The enlarged version of the field induced spectral splitting is shown in
  (b) inset. (c) and (d): The field dependence of the internal field calculated from $H_{in}=\Delta f/\gamma_n$ of the peaks denoted in (a) and (b) for both directions.
  The field dependence of the internal field for the paramagnetic state is shown by the dashed line.
}
\end{figure}

The spectral broadening below $T=6$ K results from the emergence of magnetic order. In Fig.\ref{spec2} (a) and (b), we show the spectra at $T=2$ K under
different field intensities with $B_0||b$ and $B_0\perp b$-axis. By increasing the field intensity, the spectra are broadened and a slight line splitting for the peak
from the [HCOO]$^-$ anions noted by the gray line is observed for $B_0>3.7$ T, which is more clear for $B_0||b$ (See Fig.\ref{spec2} (a) and (b)). We plot
the internal field calculated from the frequency shift $f-\gamma_n B_0$ versus field in Fig.\ref{spec2} (c) and (d). For the line shift and broadening with a
paramagnetic origination, the internal field is proportional to the magnetic field, which is shown by the dashed line. However, this is obviously not the actual situation
observed in this sample. The field dependence gives an intercept to a nonzero internal field at $B_0=0$, which is strong evidence for the magnetically  ordered
ground state.

The magnetic structure with a $G$-type antiferromagnetism with spin-canting is suggested from the spectra.
At $T=2$ K well below $T_N$ with the applied field less than 3.7 Tesla, the main characteristic of the spectrum is maintained as compared with that in the paramagnetic
phase (data not shown), and no additional line splitting due to the setup of the antiferromagnetic order is observed. This is very different from the ordinary
antiferromagnetic order\cite{Ma_PRB_90_144502}. Two scenarios can be proposed: One is the very small magnetic moment; the other one is the cancelation of the
hyperfine field on the $^1$H sites from the magnetic moments located at the two nearest Cu$^{2+}$ neighbours. At $B_0=1.5$ T, the studied peak at $T=2$ K is broadened
by $\sim 50$ kHz as compared with that at $T=5$ K well above $T_N$. By the moderate hyperfine coupling constant discussed above, the upper limit of the ordered moment
is estimated to be 0.0094$\mu_B$. In the nearly ideal quasi-1D spin chain Sr$_2$CuO$_3$, the ordered moment is $\sim0.06\mu_B$, which is reduced by strong quantum fluctuations. According to the mean-field theory\cite{Schulz_PRL_77_2790}, the ordered moment is proportional to $\sqrt{J_{\perp}/J_{//}}$. Thus, for the present Cu-MOF located at the 1D-3D crossover regime, the 0.0094$\mu_B$ ordered moment size is obviously too small to be reasonable. Thus, our results point to the second one.

Every proton in the [HCOO]$^-$ anions has two nearest Cu$^{2+}$ neighbours along all the three crystalline directions. The angle of Cu-H$_1$-Cu and Cu-H$_4$-Cu is $\sim161^o$ and $\sim155^o$, far from the exact geometric center between the two adjacent magnetic sites. The Jahn-Teller distortion again lowers the crystalline
symmetry\cite{Pato_JMCC_4_11164}. Thus, the cancelation of the internal field occurs only in ordered states with a $G$-type configuration, where the magnetic moments order antiferromagnetically for all the three directions. However, this type of antiferromagnetic order contradicts with the ferromagnetic interactions in the $ac$-plane as discussed above according to the GKA rules\cite{Goodenough_magnetism}. The discrepancy may be related with the emergence of complex DM interactions, and further identification of the antiferromagnetic order and determination of the magnetic interaction strength is needed.

The increasing Knight shifts at low temperatures indicate that the upturn behavior of the dc susceptibility results from intrinsic enhanced spin susceptibility instead of the trivial impurity effect. The $^1$H peaks shift further with respect to the Larmor frequency even in the antiferromagneic ordered state, which is
distinct with the ordinary antiferromagnets. In Cu-MOF, the (CH$_3$NH$_3$)$^+$ cations sitting on the A sites lack the spatial-inversion symmetry, resulting in the DM interaction between magnetic sites, which is important for the appearence of multiferroic behavior in its counterpart CH$_3$NH$_3$Co(HCOO)$_3$\cite{Zapf_JACS_138_1122}.
Thus, the $^1$H sites only pick the component of hyperfine field resulting from canted static moment, while that from the antiferromagnetic arrangement of moments is canceled out.

The enhancement of the internal field and line splitting due to applied magnetic field reflect the tilting magnetic moment and spin reorientations.
With $B_0//b$, the internal field shown by the original peak at low frequency increases with field strength as shown in Fig.\ref{spec2}(c), while that for $B_0\perp b$ remains nearly constant. The anisotropy again demonstrates the quasi-1D magnetic behavior induced by strong axial anisotropic magnetic interactions.
This is consistent with the field dependence of the magnetization at $T=2$ K.

\begin{figure}
\includegraphics[width=8.5cm, height=11cm]{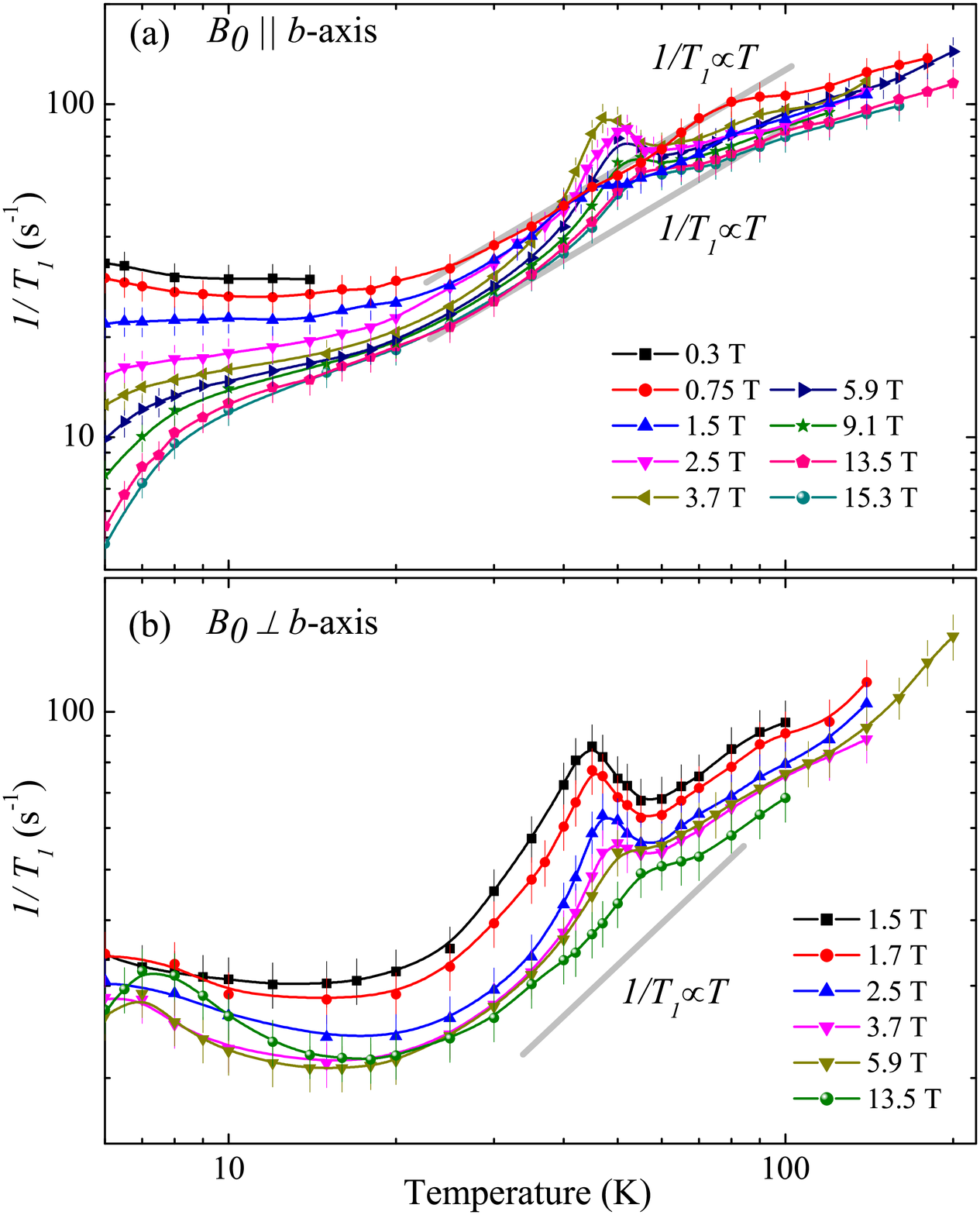}
\caption{\label{slrr1}(color online)
  The spin-lattice relaxation rate $1/T_1$ in the paramagnetic state ($T\geq7$ K)as a function of temperature under different field intensities for $B_0||b$ (a) and $B_0\perp b$-axis (b). Solid lines show the linear temperature dependence of $1/T_1$.
}
\end{figure}

Next, we concentrate on the spin excitation and its evolution under magnetic field. The spin-lattice relaxation rate $1/T_1$ is a measure of the dynamaic
spin susceptibility, whose temperature and field dependence reveal the nature of spin fluctuations in the strongly-correlated systems. In Fig.\ref{slrr1},
we show the temperature dependence of $1/T_1$ above $T_N$ at various magnetic field intesities for both $B_0||b$ (a)and $B_0\perp b$-axis (b). For $B_0=0.75$ T$||b$-axis, the $1/T_1$ shows a nearly temperature-independent behavior from $T_N$ to $T\sim20$ K, which is followed by the linear temperature dependence denoted by the gray line at higher temperatures. With the field intensity increased, besides the temperature dependence of $1/T_1$ described above, a "hump" structure is observed at temperatures ranged from $T=40$ K to 60 K. The intensity of the "hump" is enhanced firstly, reaches its maximum at $B_0=3.7$ T, and is suppressed with further increased field intensity. For $B_0\perp b$-axis, similar $1/T_1$ behavior is observed (Fig.\ref{slrr1}(b)).

The $1/T_1$ behavior shown in Fig.\ref{slrr1}(a) can be completely understood for this quasi-1D HAFC system.
For the applied magnetic field along $z$ direction, the $1/T_1$ can be expressed as
$$1/T_1=\gamma_n^2[<(\mu_0h_x)^2>+<(\mu_0h_y)^2>]\times\frac{\tau_c}{1+\omega_L^2\tau_c^2}$$,
where the the first part before "$\times$" describes the spin fluctuations at both the directions
perpendicular to the applied field\cite{Slichter_NMR}. The second part, where $\tau_c$ and $\omega_L$
denote the typical time scale of the electron system and the nuclear Larmor frequency respectively,
undergoes a maximum for $\tau_c=1/\omega_L$. Thus, we attribute the "hump" structure to the slow dynamics of the spin system.

The scaling theory for the $S=1/2$ HAFC\cite{Sachdev_PRB_50_13006} predicts that the staggered spin susceptibility
contributed by spinon excitations dominates the $1/T_1$ at low temperatures ($T\ll J/k_B$), leading to a constant $1/T_1$.
While, the uniform spin susceptibility dominating the $1/T_1$ at higher temperatures ($T<J/k_B$), results in a linear
temperature dependence of $1/T_1$. Thus, our data fit quite precisely with this theory, and provide strong evidence for the
spinon excitations at low temperatures, and again demonstrate the 1D character of this molecule magnet\cite{Takigawa_PRL_76_4612}.
This result is obviously very different with the classical spin chains, where the $T^{-3/2}$ temperature dependence of $1/T_1$ is
theoretically predicated\cite{Fisher_AJP_32_343}.

\begin{figure}
\includegraphics[width=8.5cm, height=8cm]{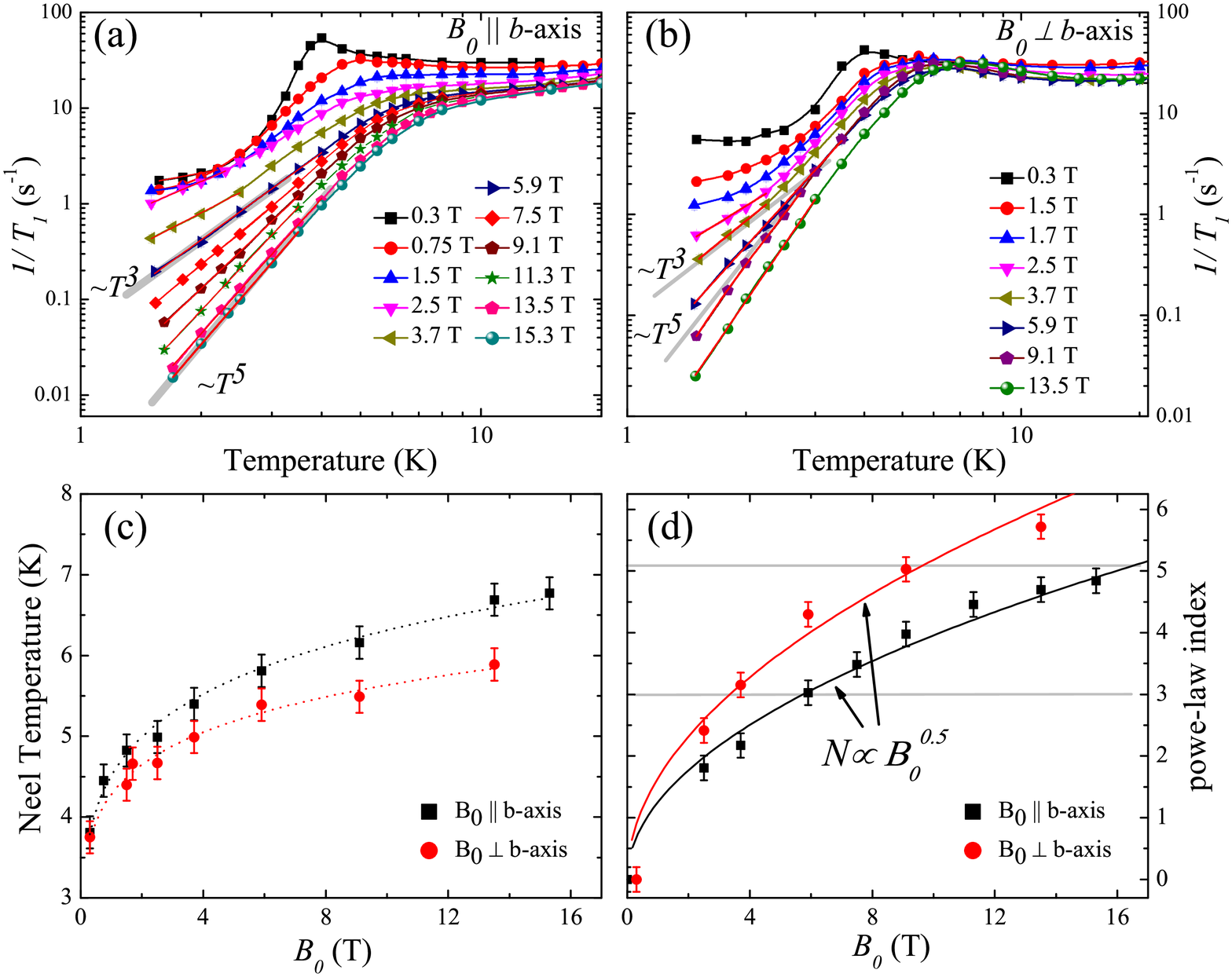}
\caption{\label{slrr2}(color online)
  The spin-lattice relaxation rate $1/T_1$ below $T=20$ K, as a function of temperature under different field intensities for $B_0||b$ (a) and $B_0\perp b$-axis (b). Solid lines show the $T^3$ and $T^5$ temperature dependence of $1/T_1$. Red straight lines are power-law fits to the $1/T_1$. (c): The field dependence of the N\'{e}el temperature. The dotted lines are guides to the eye. (d): The power-law index $N$ as a function of magnetic field. Solid lines are fits to the function $N=A\times \sqrt{B_0}$, where $A$ is the fitting parameter.
}
\end{figure}

The inter-chain coupling results in the 3D N\'{e}el order below $T_N$. In Fig.\ref{slrr2} (a) and (b), we show the temperature dependence of $1/T_1$ below $T_N$ for both field directions to study the field dependence of the spin excitations in the ordered state. For $B_0=0.3$ T, the nearly-constant $1/T_1$ decreases
sharply below $T_N$, and begins to flatten out below $T\sim3$ K for both field directions. With increasing the field intensity, we observe a power-law temperature
dependence of the $1/T_1$ for temperatures well below $T_N$, and further plot the power-law index versus field intensity in Fig.\ref{slrr2} (d).
For a conventional 3D antiferromagnet, the $1/T_1$ is mainly contributed by the scattering of nuclear spins by magnons.
Theoretical calculations\cite{Beeman_PR_166_359} predict a $T^3$-dependence of $1/T_1$ for the two-magnon Raman process,
and a $T^5$-dependence of $1/T_1$ for the three-magnon process for $T\gg\Delta$ ($\Delta$ is the excitation gap of magnons).

The evolution of the low-energy spin excitations observed in $1/T_1$ under magnetic field should be related with the enhanced effective local transverse staggered field raised by the DM interaction with the increasing field. In the well-known HAFC systems without DM-interactions such as KCuF$_3$, BaCu$_2$Si$_2$O$_7$, Sr$_2$CuO$_3$ and SrCuO$_2$, the low-lying spin excitations for the magnetically ordered state are dominated by the spin waves, and free spinons are observed at a higher energy region with a finite gap as seen from Raman scattering and neutron scattering studies\cite{Tennant_PRL_70_4003, Misochko_PRB_53_R14733,Zheludev_PRB_65_014402}. According to the mean-field theory, the spinon excitation gap size $\Delta$ is proportional to $h^{2/3}$, where $h$ denotes the effective staggered field\cite{Schulz_PRL_77_2790}, with the intensity determined by the inter-chain coupling and the orientation along the antiferromagnetic arrangement of the moment.

For the HAFC systems with strong DM interactions and very weak inter-chain coupling, the effective staggered field can be generated by the DM interaction. In copper benzoate Cu(C$_6$H$_5$COO)$_2\cdot$3H$_2$O, quasi-1D magnetic behavior is observed from susceptibility, ESR and NMR measurements\cite{Date_PTPS_46_194}, where the inter-chain coupling is very weak. The effective transverse staggered field is generated by the combined effect of anisotropic $g$-tensor and DM interactions. From theoretical results\cite{Oshikawa_PRL_79_2883, Affleck_PRB_60_1038, Xi_PRB_84_134407}, the HAFC with DM interactions under magnetic field can be mapped to simple Heisenberg spin chains with an effective staggered field. The intensity of staggered field is proportional to the applied external field and the direction is determined by $\overrightarrow{B_0}\times\overrightarrow{D}$ ($\overrightarrow{D}$ is the DM $\overrightarrow{D}$vector). This theory is evident to be successful in describing the field induced spin wave excitation gap in copper benzoate and copper pyrimidine dinitrate\cite{Dender_PRL_79_1750,Zvyagin_PRB_83_060409}.

In the present Cu-MOF sample, the DM interactions due to lack of inversion symmetry center is significant compared with a moderate inter-chain coupling ($J_{\perp}\sim0.13$ meV). In this case, the effective staggered field $h$ creating the linear attracting potential is vector sum of that raised by the inter-chain coupling and DM interaction. With the increasing field intensity, the spin-canting effect induced by the DM interaction is strongly enhanced, which is observed from the spectral analysis shown above. The staggered field created by the DM-interaction is also enhanced by increasing the magnetic field ($|\overrightarrow{h}_{DM}|\propto |\overrightarrow{B_0}\times\overrightarrow{D}|$), leading to the increasing of the attracting energy potential. The increased N\'{e}el temperature under magnetic field (See Fig.\ref{slrr2} (c)) is also consistent with this suppression of the quantum fluctuations.  The crossover from the dominating two-magnon Raman process to three-magnon scattering contributions to the nuclear relaxation may be also related with the magnetic field enhancement of spin-canting effect. At the strong magnetic field side, we fail to observe the field induced spin wave excitation gap as what happens in copper benzoate where the inter-chain coupling is very weak. Based on the Monte Carlo simulations\cite{Xi_PRB_84_134407}, the threshold for the field strength sufficient to open the energy gap in the spin wave excitations increased mildly with the inter-chain coupling. For the present sample lying in the 1D-3D crossover regime with $J_{\perp}/J\sim0.02$, the magnetic field threshold should be much higher than $B_0=15.3$ T. Thus, the field induced energy gap of the spin wave excitation is absent in our sample under the present field range. Up to now, we still do not know the physical origin for the $(B_0)^{0.5}$ field dependence of the power-law index and further theoretical calculation is needed.

In quasi-1D quantum spin systems, the magnetic field can be an effective tuning parameter to approach the quantum critical point. In the spin chain system, NiCl$_2$-4SC(NH$_2$)$_2$ (DTN) and the spin ladder compound CuBr$_4$(C$_5$H$_{12}$N)$_2$ (BPCB)\cite{Mukhopadhyay_PRL_109_177206,Klanjsek_PRL_101_137207}, the spin excitations at zero field are dominated by magnons with a finite gap, resulting from the single-ion anisotropy and strongly coupled rung. With applied magnetic field, the energy gap between spin-singlets and triplets is tuned linearly to zero due to the competing Zeeman energy in the Hamiltonian. The quantum critical point is approached at $B_{c}$. The dominating spin excitations evolve from gapped magnons to spinons in the gapless Luttinger liquid state as seen from NMR studies. Comparatively, we have realized the magnetic field suppression of quantum excitations in the low temperature ordered state in a quasi-1D spin chain system with significant DM-interactions in this study.

To conclude, we have carried out NMR study on a quasi-1D $S=1/2$ Heisenberg antiferromagnetic chain. The antiferromagnetic long range order is identified from
the spectral analysis, and a $G$-type antiferromagnetism with a ferromagnetic component is proposed. Above $T_N$, the spinon excitation is observed from the
constant $1/T_1$ at low temperatures contributed from the staggered spin susceptibility. At low temperatures well below $T_N$, the spin lattice relaxation rate $1/T_1$ is observed to flatten out toward zero temperature. By the applied field, $1/T_1$ gradually shows a power-law temperature dependence with
the index enhanced from zero to $\sim3$, and finally $\sim5$, which are the typical character for the dominating magnon scattering of the nuclear spins in
the conventional 3D classical magnet. The result supplies strong evidence for the magnetic field suppression of quantum magnetic excitations in the magnetically ordered state of the quasi-1D Heisenberg spin chain system. The mechanism for this phenomena is proposed to be related to the enhanced local transverse staggered field generated by the DM interactions under magnetic field.

This research was supported by the National Key Research and Development Program of China (Grant No. 2016YFA0401802 and 2017YFA0303201), the National Natural Science Foundation of China (Grants No. 11504377, 11874057, 11574288, U1732273, U1532153, 11774352, 21301178 and 11674331), 100 Talents Programme of Chinese Academy of Sciences (CAS) and the Major Program of Development Foundation of Hefei Center for Physical Science and Technology (Grant No. 2016FXZY001).

\end{document}